\documentclass[12pt]{iopart}
\input iopams.sty
\input epsf
\begin{document}
\jl{1}
\title[Quantum electromagnetism near a dielectric]
{The electromagnetic field near a dielectric half--space}
\author{Adam D Helfer and Andrew S I D Lang\footnote{Present address:
Department of Mathematics, Oral Roberts University, 
7777 South Lewis Avenue, Tulsa, Oklahoma 74171--0001, USA}}
\address{Department of Mathematics, University of Missouri,
         Columbia, Missouri, 65211, USA}

\begin{abstract}
We compute the expectations of the squares of the electric and magnetic
fields in the vacuum region outside a half--space filled with a uniform
non--dispersive dielectric.  This gives predictions for the
Casimir--Polder force on an atom in the `retarded' regime near a
dielectric.  We also find a positive energy density due to the
electromagnetic field.  This would lead, in the case of two parallel
dielectric half--spaces, to a positive, separation--independent
contribution to the energy density, besides the negative,
separation--dependent Casimir energy.  Rough estimates suggest that
for a very wide range of cases, perhaps including all realizable ones,
the total energy density between the half--spaces is positive.
\end{abstract}

\submitted
\maketitle
\def\kt{{k_{\rm T}}}
\def\xt{{x_{\rm T}}}
\def\EE{{E^{\rm E}}}
\def\EM{{E^{\rm M}}}
\def\BE{{B^{\rm E}}}
\def\BM{{B^{\rm M}}}
\def\ET{{E_{\rm T}}}
\def\BT{{B_{\rm T}}}

\section{Introduction}
In this paper, we investigate the quantum electromagnetic field in the
vacuum region outside a half--space filled with a uniform,
non--dispersive dielectric.  We compute the expectations of the squares
of the electric and magnetic fields in this region.  We have two
motivations for this.

First, the problem is natural in the study of quantum optics.  Indeed,
other workers have already investigated some aspects of this situation,
for example, the effects of a nearby dielectric on atomic transition
rates (see, e.g., Khosravi and Loudon 1991, 1992).  Here we compute the
expectations of the squares of the electric and magnetic fields, thus
providing predictions of the Casimir--Polder force on an (electrically
or magnetically) polarizable atom near the dielectric.  These
predictions test the ultraviolet renormalization of the the theory at a
deeper level than do the transition--rate ones.

Our main motivation, however, comes from the hypothesized uses of
negative energy densities to fuel exotic general--relativistic and
thermodynamic effects.  Serious workers have considered the possibility
that negative energy densities might give rise to ``worm holes,''
``warp drives'' and ``time machines.''  Such predictions depend on
being able to generate persistent negative energy densities.  At
present, the only way that this might be achieved within reasonably
well--understood physics is via Casimir--type effects.  In the original
Casimir (1948) effect, for example, the energy density due to the
quantum electromagnetic field between two perfect parallel plane
conductors is predicted to be negative.  It should immediately be
remarked that this negative energy density has never been directly
observed.\footnote{Laboratory experiments measure the force between the
plates, that is, th component ${\hat T}_{zz}$ of the stress--energy
(Sparnaay 1957, 1958; Lamoreaux 1997, Bordag {\it et al.} 1998).  
The energy density is ${\hat
T}_{tt}$.  These two operators do not commute.  There is a connection
between them, in that the long--time average of the force is minus the
gradient of the energy, but present experiments seem far from being
able to measure ${\hat T}_{tt}$.  This operator may as a matter of
principle not be directly observable; see Helfer 1998.} Still, it is
this prediction which has generated an enormous amount of theoretical
work, because the possible consequences are so spectacular.

We wanted to know what would happen to the prediction of negative
energy densities if the plates were no longer idealized as perfect
conductors.  A realistic treatment of this would require a theory of
the quantum electromagnetic field in inhomogeneous absorptive and
dispersive media at finite temperature.  Such theories are only
now under development (see, e.g., Matloob {\it et al.} 1995), 
so it seems wise to consider as a first
step the case of a non--absorptive, non--dispersive medium at zero
temperature.  Thus we shall consider the case of a half--space filled
with a material of (frequency--independent) dielectric constant
$\epsilon$.  While the case of a perfect conductor is formally the
limit $\epsilon\uparrow\infty$ of this, an imperfect conductor is not
well represented by such a model with $\epsilon$ finite.  So we shall
not be able to make any positive predictions about the behavior of real
conductors.

Still, our results are strong enough to bear on the case of
conductors.  We shall find that, for dielectrics, finite--$\epsilon$
effects {\em cannot} be neglected, especially in computations of the
electromagnetic contribution to the energy density.  In the case of two
parallel half--spaces, these finite--$\epsilon$ corrections do not
alter the attractive nature of the Casimir force, but may contribute a
positive, separation--independent energy density which dominates the
negative, separation--dependent, Casimir energy density.  This strongly
suggests that only after a careful treatment of the physics of real
conductors will we know whether the perfect--conductor idealization is
adequate for computing the energy density in such cases.

It is not easy to say accurately and briefly why a finite dielectric
constant should modify the energy density to this degree, because the
physics is non--local and depends on quantum interference.  The
presence of a polarizable medium in a region alters the field
operators, by causing reflection and refraction of modes at the
boundary.  If the geometry is particularly simple (a plane interface)
and the reflection sufficiently idealized (a perfect conductor), one
has a great deal of cancellation.  Small deviations from these
idealizations can potentially lead to large effects.  This is beacuse
the energy density and the squares of the field strengths are 
defined by ultraviolet--divergent integrals (and
must be renormalized).

To explain the situation more quantitativley, we first review some
aspects of the Casimir effect, and then discuss the idealizations that
have been made and how they might be expected to be modified in a more
realistic treatment.

Between two perfect parallel plane conductors, one finds that the
renormalized energy density is given by
\begin{equation}
 \langle {\hat T}_{00}\rangle _{\rm ren}
  ={1\over 2}\langle {\hat E}^2+{\hat B}^2\rangle _{\rm ren}
  =-{{\pi ^2\hbar c}\over{720 l^4}}\; ,\label{casimir}
\end{equation}
where $l$ is the distance between the plates.  That this is independent
of position can be shown on invariance grounds (and relies on the ideal,
perfect--conductor boundary conditions).  However, the electric and
magnetic fields are {\em not} position--independent; one finds
\begin{eqnarray}
 \langle {\hat E}^2\rangle _{\rm ren}
  &=&-{{\pi ^2\hbar c}\over{720 l^4}}
    +{{\pi ^2\hbar c}\over{16 l^4}} {{3-2\sin ^2 (\pi z/l)}\over
      {\sin ^4 (\pi z/l)}}\label{ahab}\\
 \langle {\hat B}^2\rangle _{\rm ren}
  &=&-{{\pi ^2\hbar c}\over{720 l^4}}
    -{{\pi ^2\hbar c}\over{16 l^4}} {{3-2\sin ^2 (\pi z/l)}\over
      {\sin ^4 (\pi z/l)}}
\end{eqnarray}
at distance $z$ from one plate.  Near one plate, as $z\downarrow 0$, we
find the asymptotic forms
\begin{eqnarray}
 \langle {\hat E}^2\rangle _{\rm ren}
  &\sim &+{{3\hbar c}\over{16 \pi ^2z^4}}\\
 \langle {\hat B}^2\rangle _{\rm ren}
  &\sim&-{{3\hbar c}\over{16 \pi ^2z^4}}\; .\label{beelzebub}
\end{eqnarray}
In other words, the renormalized expectations of ${\hat E}^2$ and ${\hat
B}^2$ both diverge near a perfectly conducting plate, but there is a
perfect cancellation between the divergent terms, leaving only a finite
result.

Several commments on this are in order.
First, the negative expectation of ${\hat B}^2$ occurs because it is a
renormalized quantity, and means that the fluctuations of $\hat B$ are
less than those of the Minkowski vacuum.
Second, the divergences of $\langle {\hat E}^2\rangle _{\rm ren}$
and $\langle {\hat B}^2\rangle _{\rm ren}$ as $z\downarrow 0$ are not
expected to be physical, but rather arise from the idealized boundary
conditions used.  A real conductor would not be well approximated by a
perfect conductor within atomic distances, and probably not within its
plasma wavelength.  Thus the expressions (\ref{ahab})--(\ref{beelzebub})
are really only expected
to be valid when one is sufficiently far from the conductor to neglect
atomic structure and finite skin--depth.

Still, one is led to ask what would happen if the antisymmetry between
the divergent parts of $\langle {\hat E}^2\rangle _{\rm ren}$
and $\langle {\hat B}^2\rangle _{\rm ren}$ could be disturbed.  Could
one produce energy densities much greater in magnitude than the Casimir
expression (\ref{casimir})?  A natural way to try to do this is to replace the
perfect conductor by a dielectric, and this is what we have done here. 
Of course, our model is not expected to be accurate within atomic
distances or even scales of the order of a skin--depth.  Still, we shall be
able to draw some interesting conclusions.

We are able to compute $\langle {\hat E}_z^2\rangle _{\rm ren}$,
$\langle {\hat E}_{\rm T}^2\rangle _{\rm ren}$,
$\langle {\hat B}_z^2\rangle _{\rm ren}$
and $\langle {\hat B}_{\rm T}^2\rangle _{\rm ren}$ explicitly, as functions of
the distance $z$ from the dielectric boundary and of the dielectric
susceptibility $\chi$.  The expressions have the form
\begin{equation}
 \eta \hbar c/z^4
\end{equation}
where the $\eta$s are transcendental functions of $\chi$. 
(See equations \ref{dagobert}{}, \ref{ephraim}{}, \ref{fandango}{}, 
\ref{gerard}{}.)  We find in particular that the energy density in the
vacuum half--space has the form $\eta _\rho \hbar c/z^4$, where $\eta _\rho$ is
a positive function of $\chi$.  This means that the total energy 
per unit surface area of the electromagnetic field on the vacuum side,
\begin{equation}
  \int _0^\infty (\eta _\rho \hbar c/z^4)\, \rmd z\, ,
\end{equation}
is divergent.  This is unphysical and again can be ascribed to the
oversimplification of our model, where all modes, of whatever frequency,
are equally affected by the dielectric.  In a more realistic model, the
dielectric's atomic structure would be taken into account.  This would
mean that at small distances (of the order of the skin depth probably
and at the atomic scale certainly) the energy density would not be given
by $\eta _\rho\hbar c/z^4$, 
but by some other, presumably finite, expression. 
Correspondingly, we ought really to think of our theory as an effective
field theory valid only up to frequencies corresponding to wavelengths
of order the skin depth or so.

In the next section, we outline the technical details of the
computations.  In Section 3, we summarize the asymptotic behaviors of
the squares of the $\eta$s for the squares of the fields, and present
the graphs of these functions.
Section 4 summarizes the behavior of the expectation of the stress
tensor.  Section 5 contains discussions of the significance of our
results, and Section 6 recapitualiates the main conclusions.

\section{The Computation}
\subsection{The Orthonormal Eigenmodes}
The case of a half--space uniformly filled with a dielectric has been
studied earlier, and we shall use the orthonormal eigenmodes as given by
Carniglia and Mandel (1971).

We shall take the $z$ axis to be normal to the interface, with $z$
increasing in the vacuum region.  We take advantage of the
translational symmetries in time and in the $x_{\rm T}=(x,y)$
directions to resolve all modes by Fourier transforms in these
variables, with Fourier transform variables $\omega$ and $k_{\rm T}$. 
These Fourier transform variables thus retain their senses on both sides
of the interface.

The dielectric constant is $\epsilon =1+\chi$.  The wave number in the
$z$--direction is $k$ in the vacuum and $\tilde k$ in the dielectric. 
Thus we have
\begin{eqnarray}
  {\tilde k}^2+k_{\rm T}^2&=&\epsilon \omega ^2 \quad\mbox{for}\quad
     z<0\quad\mbox{(dielectric)}\\
  k^2+k_{\rm T}^2&=& \omega ^2 \quad\mbox{for}\quad
     z>0\quad\mbox{(vacuum).}
\end{eqnarray}
In what follows ${\hat\kt}$ and ${\hat e}_z$ are the unit vectors
in the $\kt$ and $z$--directions.  In later sections, hats will
indicate field operators, too, but no confusion should arise.

We shall only need the modes on the vacuum side of the interface.
The transverse electric component of the `electric' field mode (where
$E$ is normal to the plane of incidence) incident from the left
($\tilde{k}>0$), is
\begin{equation}
\EE_{\tilde{k} \kt} = 
(2\epsilon )^{-1/2}
  \frac{2 \tilde{k}}{\tilde{k} + k}\rme ^{\rmi k z}
\rme ^{\rmi \kt\cdot \xt} (\hat{\kt} \times \hat{e}_z) \, .
\end{equation}
The transverse magnetic component of the `electric' field mode incident 
from the left ($\tilde{k}>0$) is
\begin{equation}
\EM_{\tilde{k} \kt} = 
(\sqrt{2}\omega )^{-1}\frac{2 \tilde{k}}{\tilde{k} +\epsilon k}
   \rme ^{\rmi k z}
\rme ^{\rmi \kt \cdot \xt}
\left(\kt \hat{e}_z-k\hat{\kt} \right)\, .
\end{equation}
The transverse electric component of the `electric' field mode incident 
from the right (${k}>0$) is
\begin{equation}
\EE_{k \kt} = 
{\sqrt{2}}^{-1}\left( \rme ^{-\rmi {k} z}+
\frac{k-\tilde{k}}{k+\tilde{k}}
  \rme ^{\rmi {k} z}\right)\rme ^{\rmi \kt \cdot \xt}
 (\hat{\kt} \times \hat{e}_z)\, .
\end{equation}
The transverse magnetic component of the `electric' field mode incident 
from the right (${k}>0$) is
\begin{equation}
\fl\EM_{{k} \kt} = 
    (\sqrt{2}\omega )^{-1} \Bigg( \rme ^{-\rmi {k} z}
       \left(\kt \hat{e}_z
       +k\hat{\kt} \right)
       +\frac{\epsilon k-\tilde{k}}{\epsilon k+\tilde{k}}
          \rme ^{\rmi {k} z}
       \left(\kt \hat{e}_z
       -k\hat{\kt} \right)
       \Bigg)\rme ^{\rmi \kt\cdot \xt}\, .
\end{equation}
The transverse electric component of the `magnetic' field mode incident 
from the left ($\tilde{k}>0$) is
\begin{equation}
\BE_{\tilde{k} \kt} = 
(2 \epsilon\omega ^2)^{-1/2}\frac{2 \tilde{k}}{\tilde{k} +k}
   \rme ^{\rmi k z}
\rme ^{\rmi \kt \cdot \xt}
\left(k\hat{\kt} -\kt\hat{e}_z\right)\, .
\end{equation}
The transverse magnetic component of the `magnetic' field mode incident 
from the left ($\tilde{k}>0$) is
\begin{equation}
\BM_{\tilde{k} \kt} = 
2^{-1/2}\frac{2 \tilde{k}}{\tilde{k} +\epsilon k}\rme ^{\rmi k z}
\rme ^{\rmi \kt \cdot \xt} (\hat{\kt} \times \hat{e}_z)\, .
\end{equation}
The transverse electric component of the `magnetic' field mode incident 
from the right (${k}>0$) is
\begin{equation}
\fl\BE_{{k} \kt} = 
-2^{-1/2}\omega ^{-1}\Bigg( \rme ^{-\rmi {k} z}
\left(k\hat{\kt} +\kt\hat{e}_z\right)
-\frac{k-\tilde{k}}{k+\tilde{k}}\rme ^{\rmi {k} z}
\left(k\hat{\kt}-\kt\hat{e}_z \right)
\Bigg)\rme ^{\rmi \kt \cdot \xt}\, .
\end{equation}
The transverse magnetic component of the `magnetic' field mode incident 
from the right (${k}>0$) is
\begin{equation}
\BM_{k \kt} = 
2^{-1/2}\left( \rme ^{-\rmi {k} z}+
\frac{\epsilon k-\tilde{k}}{\tilde{k}+\epsilon k}\rme ^{\rmi {k} z}\right)
\rme ^{\rmi \kt\cdot\xt}
 (\hat{\kt} \times \hat{e}_z)\, .
\end{equation}

The electric and magnetic field operators are thus given by
\begin{eqnarray}
\fl\hat{E}({\bf x},t)=\frac{1}{(2 \pi)^3}\int_{\tilde{k}>0}\rmd ^3\tilde{k}
   \sum_{\lambda={\rm E},{\rm M}}
   \sqrt{\omega}\left(\hat{a}^{\lambda}_{\tilde{k}\kt}
   E^{\lambda}_{\tilde{k}\kt}\rme ^{-\rmi \omega t} + \mbox{h.c.}\right)
        \nonumber\\
 +\frac{1}{(2 \pi)^3}\int_{{k}>0}\rmd ^3{k}
\sum_{\lambda={\rm E},{\rm M}}\sqrt{\omega}\left(\hat{a}^{\lambda}_{{k}\kt}
E^{\lambda}_{{k}\kt}\rme ^{-\rmi \omega t} +\mbox{h.c.}\right)
\end{eqnarray}
and
\begin{eqnarray}
\fl\hat{B}({\bf x},t)=\frac{1}{(2 \pi)^3}\int_{\tilde{k}>0} \rmd ^3\tilde{k}
\sum_{\lambda={\rm E},{\rm M}}
   \sqrt{\omega}\left(\hat{a}^{\lambda}_{\tilde{k}\kt}
B^{\lambda}_{\tilde{k}\kt}\rme ^{-\rmi \omega t} +\mbox{h.c.}\right)
             \nonumber\\
+\frac{1}{(2 \pi)^3}\int_{{k}>0}\rmd ^3{k}
\sum_{\lambda={\rm E},{\rm M}}\sqrt{\omega}\left(\hat{a}^{\lambda}_{{k}\kt}
B^{\lambda}_{{k}\kt}\rme ^{-\rmi \omega t} +\mbox{h.c.}\right)\, ,
\end{eqnarray}
where the creation and annihilation 
operators satisfy the commutation relations
\begin{equation}
\left[\hat{a}^{\lambda^{\prime}}_{\tilde{k}^{\prime}\kt ^{\prime}}, 
\hat{a}^{\lambda^{*}}_{\tilde{k} \kt}\right]=
4 \pi^3 \hbar \delta_{\lambda \lambda^{\prime}} 
\delta(\tilde{k}-\tilde{k}^{\prime})
\delta(\kt -\kt ^{\prime}),
\end{equation}
and
\begin{equation}
\left[\hat{a}^{\lambda^{\prime}}_{k^{\prime}\kt ^{\prime}}, 
\hat{a}^{\lambda^{*}}_{k \kt}\right]=
4 \pi^3 \hbar \delta_{\lambda \lambda^{\prime}} \delta(k-k^{\prime})
\delta(\kt -\kt ^{\prime})\, .
\end{equation}

\subsection{Computation of $\hat{E}_z^2$}
We shall outline the computation of $\hat{E}_z^2$.  Computations of the
squares of the other field components follow the same pattern.

We use a standard point--splitting in imaginary time, and set $\rmi\tau
=t'-t$.  Then we have
\begin{eqnarray}
\fl\langle \hat{E}^2_z \rangle = \frac{1}{(2 \pi)^3}\int_{\tilde{k}>0} 
\rmd ^3\tilde{k}
\frac{\kt ^2}{2 \omega}\left(\frac{2 \tilde{k}}{\tilde{k} + 
\epsilon k}\right)
\left(\frac{2 \tilde{k}}{\tilde{k} + \epsilon k}\right)^{*}
\rme ^{\rmi (k-k^{*})z}\rme ^{- \omega \tau}\nonumber\\
+\frac{1}{(2 \pi)^3}\int_{{k}>0}\rmd ^3{k}
\frac{\kt ^2}{2 \omega}
\left(\rme ^{-\rmi k z}+\frac{\epsilon k - \tilde{k}}{\epsilon k+\tilde{k}}
\rme ^{\rmi k z}\right)
\left(\rme ^{-\rmi k z}+\frac{\epsilon k - \tilde{k}}{\epsilon k+\tilde{k}}
\rme ^{\rmi k z}\right)^{*}\rme ^{- \omega \tau}\, .
\end{eqnarray}
We rewrite the integral over $\tilde{k}>0$ as an integral over $k>0$ 
(representing plane waves) plus an integral over 
$0<\kappa<\omega \sqrt{\chi}$ (representing evanescent waves):
\begin{eqnarray}
\fl\langle \hat{E}^2_z \rangle = \frac{1}{(2 \pi)^2}\int_0^{\infty}\rmd \omega 
\int_0^{\omega} \rmd k (\omega^2-k^2)\left(1+
\frac{\epsilon k -\tilde{k}}{\epsilon k +\tilde{k}}\cos 2 k z\right)
\rme ^{- \omega \tau}\nonumber\\
+ \frac{1}{(2 \pi)^2}\int_0^{\infty}\rmd \omega 
\int_0^{\omega \sqrt{\chi}} \rmd \kappa (\omega^2 + \kappa^2)
\frac{2 \epsilon \kappa \tilde{k}}{\tilde{k}^2+\epsilon^2 \kappa^2}
\rme ^{-2 \kappa z}\rme ^{- \omega \tau}\, ,
\end{eqnarray}
where we have used the relationships
\begin{equation}
\kt ^2=\left\{
\begin{array}{cc}
\omega^2 + \kappa^2 &\mbox{for} \quad \tilde{k}<\omega \sqrt{\chi}\\
\omega^2 - k^2 &\mbox{for} \quad \tilde{k}>\omega \sqrt{\chi}
\end{array}\right.
\end{equation}
and have performed the simple polar angle integration.

Changing variables $k = \omega \xi$ in the first integral and $\kappa = 
\omega \sqrt{\chi} \xi$ in the second integral we obtain
\begin{eqnarray}
\fl\langle \hat{E}^2_z \rangle =  \frac{1}{(2 \pi)^2}\int_0^{\infty}\rmd \omega 
\int_0^1 \rmd \xi \,\omega^3 (1-\xi^2)\left(1+
\frac{\epsilon \xi -\sqrt{\chi + \xi^2}}
{\epsilon \xi +\sqrt{\chi + \xi^2}}\cos 2 \omega \xi z\right)
\rme ^{- \omega \tau}\nonumber\\
+  \frac{1}{(2 \pi)^2}\int_0^{\infty}\rmd \omega 
\int_0^1 \rmd \xi \,\omega^3 (1 + \sqrt{\chi} \xi^2)
\frac{2 \epsilon \xi \sqrt{\chi + \xi^2}}{1+(\epsilon^2-1)\xi^2}
\rme ^{-2 \omega \sqrt{\chi} \xi z}\rme ^{- \omega \tau}\, .
\end{eqnarray}
Integrating over $\omega$ gives
\begin{eqnarray}
\fl\langle \hat{E}^2_z \rangle =  \frac{\hbar c}{(2 \pi)^2}
\int_0^1\left[ 6\frac{1-\xi^2}{\tau^4}+6(1-\xi^2) 
\frac{\epsilon \xi -\sqrt{\chi + \xi^2}}
{\epsilon \xi +\sqrt{\chi + \xi^2}}\frac{16 z^4 \xi^4 
  - 24 z^2 \xi^2 \tau^2 
+\tau^4}{(4 z^2 \xi^2 + \tau^2)^4}\right.\nonumber\\
\left. + 
\frac{12 \epsilon \sqrt{\chi} \xi (1+\chi \xi^2)\sqrt{1-\xi^2}}
{(1+(\epsilon^2-1)\xi^2)(2 z \sqrt{\chi} \xi+\tau)^4}\right] \rmd \xi\, .
\end{eqnarray}
Thus we have reduced the problem of finding the expectation value of the 
square of the $z$--component of the electric field to a one--dimensional 
integral.  The integral can be evaluated using contour integration in the 
complex plane and by exploiting Cauchy's residue theorem.  After 
integrating and extensive algebra we obtain the following formally 
divergent expression for the expectation value:
\begin{eqnarray}
\fl\langle \hat{E}^2_z \rangle =  
\lim_{\tau \rightarrow 0} \Bigg( \frac{\hbar c}{\pi^2 \tau^4} 
+ \frac{\hbar c}{(2 \pi)^2 z^4}\Bigg[
\frac{1}{16 \chi^{3/2}}\Bigg(
2 \sqrt{\chi}(6 \epsilon^2 - 3 \epsilon^{3/2}-2 \chi)
             \nonumber\\
 + 6 \epsilon(1-2 \epsilon^2 + 2 \chi)\ln (\sqrt{\epsilon}+\sqrt{\chi})
             \nonumber\\
+\frac{6 \epsilon^2(\epsilon^2-\chi-1)}{\sqrt{\epsilon^2-1}}
\ln \left(\frac{\sqrt{\epsilon+1}-1}{\sqrt{\epsilon+1}+1}
(\sqrt{\epsilon+1}+\sqrt{\epsilon})^2
\right)\Bigg)\Bigg]+ O(\tau)\Bigg)\, .
\end{eqnarray}
Subtracting the (again divergent) vacuum (Minkowski space) expectation value
$\langle \hat{E}^2_z \rangle{\rm Minkowski} =  
\lim_{\tau \rightarrow 0} \hbar c/{\pi^2 \tau^4}$
and taking the limit as $\tau \rightarrow 0$ gives 
the exact renormalized expectation value:
\begin{eqnarray}
\fl\langle \hat{E}^2_z \rangle_{\rm ren} =  
 \frac{\hbar c}{(2 \pi)^2 z^4}\Bigg[
\frac{1}{16 \chi^{3/2}}\Bigg(
2 \sqrt{\chi}(6 \epsilon^2 - 3 \epsilon^{3/2}-2 \chi)\nonumber\\
+ 6 \epsilon(1-2 \epsilon^2 + 2 \chi) 
\ln (\sqrt{\epsilon}+\sqrt{\chi})\nonumber\\
+\frac{6 \epsilon^2(\epsilon^2-\chi-1)}{\sqrt{\epsilon^2-1}}
\ln \left(\frac{\sqrt{\epsilon+1}-1}{\sqrt{\epsilon+1}+1}
(\sqrt{\epsilon+1}+\sqrt{\epsilon})^2
\right)\Bigg)\Bigg]\label{dagobert}\\
\lo= \frac{\hbar c \eta_z^E }{z^4}\, ,
\end{eqnarray}
say,
where the coefficient $\eta_z^E$ is a function of $\chi$.

The renormalized expectations of the squares of the other components can
be calculated by the same techniques.  They are given by:
\begin{eqnarray}
\fl\langle \hat{E}^2_{\rm T} \rangle_{\rm ren} = 
\frac{\hbar c}{(2 \pi)^2 z^4}\Bigg[
\frac{1}{16 \chi^{3/2}}\Bigg(
2 \sqrt{\chi}(6 - 3 \sqrt{\epsilon}-2 \chi)
- 6(1-2 \epsilon \chi)\nonumber\\
\cdot \ln (\sqrt{\epsilon}+\sqrt{\chi})
-\frac{6 \epsilon^2 \chi}{\sqrt{\epsilon^2-1}}
\ln \left(\frac{\sqrt{\epsilon+1}-1}{\sqrt{\epsilon+1}+1}
(\sqrt{\epsilon+1}+\sqrt{\epsilon})^2
\right)\Bigg)\Bigg]\label{ephraim}\\
\lo= \frac{\hbar c \eta_{\rm T}^E }{z^4}\, ;
\end{eqnarray}
\begin{eqnarray}
\fl\langle \hat{B}^2_z \rangle_{\rm ren} = 
\frac{\hbar c}{(2 \pi)^2 z^4}\Bigg[
\frac{1}{16 \chi^{3/2}}\Big(
2 \sqrt{\chi}(12 - 9 \sqrt{\epsilon}-2 \chi)\nonumber\\
 - 6(1-2 \chi) \ln \left(\sqrt{\epsilon}+\sqrt{\chi}\Big)
\right)\Bigg]\label{fandango}\\
\lo= \frac{\hbar c \eta_z^B }{z^4}\, ;
\end{eqnarray}
and
\begin{eqnarray}
\fl\langle \hat{B}^2_{\rm T} \rangle_{\rm ren} =  
\frac{\hbar c}{(2 \pi)^2 z^4}\Bigg[
\frac{1}{16 \chi^{3/2}}\Bigg(
2 \sqrt{\chi}(6 +6 \epsilon^2 -2 \chi -3(\epsilon+2)\sqrt{\epsilon})
         \nonumber\\
+ 6(\epsilon-2 \epsilon^3+2 \chi)
\ln (\sqrt{\epsilon}+\sqrt{\chi})
          \nonumber\\
+6 \epsilon^2 \sqrt{\epsilon^2-1}
\ln \left(\frac{\sqrt{\epsilon+1}-1}{\sqrt{\epsilon+1}+1}
(\sqrt{\epsilon+1}+\sqrt{\epsilon})^2
\right)\Bigg)\Bigg]\label{gerard}\\
\lo= \frac{\hbar c \eta_{\rm T}^BB }{z^4}\, .
\end{eqnarray}

These expressions are very complicated, and the characters of the
functions $\eta (\chi )$ will be investigated in the next section.  For
the present, we remark that the sucessful renormalization provides a
very strong check on the computations, since the term of order $\tau
^{-4}$ must cancel perfectly against the term from Minkowski space, and
the remaining potential poles in $\tau$ (of orders
$\tau ^{-3}$, $\tau ^{-2}$ and $\tau ^{-1}$) must
vanish identically.  Another check is provided by the vanishing of
$\langle {\hat T}_{zz}\rangle_{\rm ren}$, as will be discussed in
Section 4.

\section{The Squares of the Fields}
In the previous section, we found the expectations of the squares of the
fields explicitly.  In each case the result had the form $\hbar c\eta
/z^4$, where $z$ was the distance to the interface and $\eta$ was a
complicated transcendental function of the susceptibility $\chi$.  In
this section, we present the graphs of the functions $\eta$, as well as
their limiting behaviors for $\chi\downarrow 0$ and $\chi\uparrow\infty$.

The functions $\eta ^{E,B}_{z,{\rm T}}$ (scaled to have a common
limiting value) are presented in figure~1.   They are in each case
monotonic and approach a constant value asymptotically, but the approach
is extremely slow.

For $\chi \downarrow 0$, we have
\begin{eqnarray}
\langle \hat{E}^2_z \rangle_{\rm ren} 
  = \frac{\hbar c}{(2 \pi)^2} \frac{9}{80 z^4} \chi + O(\chi^2)\, ,\\
\langle \hat{E}^2_{\rm T} \rangle_{\rm ren} 
     = \frac{\hbar c}{(2 \pi)^2} \frac{7}{40 z^4} \chi + O(\chi^2)\, \\
\langle \hat{B}^2_z \rangle_{\rm ren} 
  = -\frac{\hbar c}{(2 \pi)^2} \frac{1}{80 z^4} \chi + O(\chi^2)\, ,\\
\langle \hat{B}^2_{\rm T} \rangle_{\rm ren} 
  = -\frac{\hbar c}{(2 \pi)^2} \frac{3}{40 z^4} \chi + O(\chi^2)\, .
\end{eqnarray}
It is good to observe that all the above expectation values tend to zero as 
$\chi \downarrow 0$ (vacuum).  

For $\epsilon \gg 1$ (that is, $\chi\uparrow\infty$), we find
\begin{eqnarray}
\langle \hat{E}^2_z \rangle_{\rm ren} = \frac{\hbar c}{16 \pi^2 z^4}
- \frac{\hbar c}{(2 \pi)^2} \frac{3}{16 z^4}\frac{1}{\sqrt{\epsilon}} 
+ O(1 / \epsilon)\, ,\\
\langle \hat{E}^2_{\rm T}\rangle_{\rm ren} = \frac{\hbar c}{8 \pi^2 z^4}
- \frac{\hbar c}{(2 \pi)^2} \frac{3}{4 z^4}\frac{1}{\sqrt{\epsilon}} 
+ O(1 / \epsilon)\, ,\\
\langle \hat{B}^2_z \rangle_{\rm ren} = -\frac{\hbar c}{16 \pi^2 z^4}
+ \frac{\hbar c}{(2 \pi)^2} \frac{3}{8 z^4}\frac{\ln \epsilon}
{\sqrt{\epsilon}} + O(1 / \sqrt{\epsilon})\, ,\\
\langle \hat{B}^2_{\rm T} \rangle_{\rm ren} = -\frac{\hbar c}{8 \pi^2 z^4}
+ \frac{\hbar c}{(2 \pi)^2} \frac{3}{8 z^4}\frac{\ln \epsilon}
{\sqrt{\epsilon}} + O(1 / \sqrt{\epsilon})\, .
\end{eqnarray}
We note that in the limit $\chi\uparrow\infty$, these quantities attain
the values they would have in the case of the half--space outside a
perfectly conducting plane (see {\it e.g.} Barton 1990).  
This is in accord with the usual
formal identification of perfect conductors with dielectrics of
infinite susceptibility.  However, the approach to this limit is rather
slow.  One needs $\chi\approx 102$ for $\eta ^B_z$ to be within 50\% of
its limiting value, and $\chi\approx 14\,400$ to be within 10\%.

\begin{figure}
\begin{center}
\leavevmode
\epsfysize=3in 
\epsfbox{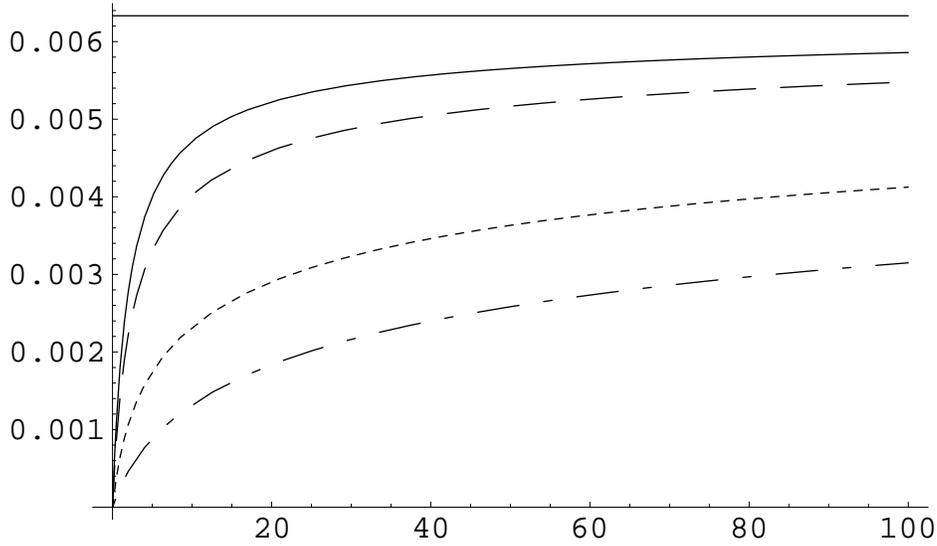}
\end{center}
\caption{The dependences of the squares of the fields on the
susceptibility $\chi$ (the abscissa).  The horizontal line 
at the top is the common asymptote. 
Below that, in descending order, are $\eta ^E_z$, then
$(1/2)\eta ^E_{\rm T}$, then $-(1/2)\eta ^B_{\rm T}$, and
finally $-\eta ^B_z$.}
\end{figure}

For completeness, we list the limiting behaviors of squares of the full
fields:
\begin{eqnarray}
\langle \hat{E}^2 \rangle_{\rm ren} =
\frac{\hbar c}{(2 \pi)^2 z^4}\frac{23}{80}\chi+O(\chi^2)\\
\langle \hat{B}^2 \rangle_{\rm ren} =
-\frac{\hbar c}{(2 \pi)^2 z^4}\frac{7}{80}\chi+O(\chi^2)
\end{eqnarray}
for $\chi\downarrow 0$, and
\begin{eqnarray}
\langle \hat{E}^2 \rangle_{\rm ren} =
\frac{3 \hbar c}{16 \pi^2 z^4} -\frac{\hbar c}{(2 \pi)^2 z^4}\frac{23}{80}
\frac{1}{\sqrt{\epsilon}}+O(1/ \epsilon)\\
\langle \hat{B}^2 \rangle_{\rm ren} =
-\frac{3 \hbar c}{16 \pi^2 z^4} +\frac{\hbar c}{(2 \pi)^2 z^4}\frac{3}{4}
\frac{\ln \epsilon}{\sqrt{\epsilon}}+O(1/ \sqrt{\epsilon})
\end{eqnarray}
for $\chi\uparrow\infty$.

\section{The Stress Tensor}
By symmetry considerations, the expectation
of the renormalized stress tensor must be diagonal.  
The `$z$' component of the divergence constraint implies that
$\langle\hat{T}_{zz}\rangle _{\rm ren}$ must be independent of $z$;
however, as all components must be multiples of $1/z^4$, this component
must be zero.  (The verification that one does get zero using our values
of the squares of the fields provides another check on our calculation.)
Since the tensor is trace--free and the `$xx$' and `$yy$' components
must be equal, there is only one algebraically independent component.
We may take this to be the energy density $\rho
=\langle {\hat T}_{tt}\rangle _{\rm ren}$, the other non--zero terms
being $\langle {\hat T}_{xx}\rangle _{\rm ren}
=\langle {\hat T}_{yy}\rangle _{\rm ren}=\rho /(2c^2)$.

The renormalized energy density 
is given by
\begin{eqnarray}
\fl\langle\hat{T}_{00} \rangle_{\rm ren} 
  = \frac{\hbar c}{(2 \pi)^2 z^4}\Bigg[
\frac{1}{16 \chi^{3/2}}\Bigg(
2 \sqrt{\chi}(6 \epsilon^2 +12 - 4 \chi -3(\epsilon+3)\sqrt{\epsilon})
   \nonumber\\
+  6 (\epsilon-1-2 \epsilon^3 + 2(\epsilon+1) \chi)
\ln (\sqrt{\epsilon}+\sqrt{\chi})
   \nonumber\\
+ \frac{6 \epsilon^2(\epsilon^2-\chi-1)}{\sqrt{\epsilon^2-1}}
\ln \left(\frac{\sqrt{\epsilon+1}-1}{\sqrt{\epsilon+1}+1}
(\sqrt{\epsilon+1}+\sqrt{\epsilon})^2
\right)\Bigg)\Bigg]\label{ichabod}\\
\lo= \frac{\hbar c \eta _\rho}{z^4}\, ,
\end{eqnarray}
say.

\begin{figure}
\begin{center}
\leavevmode
\epsfysize=3in \epsfbox{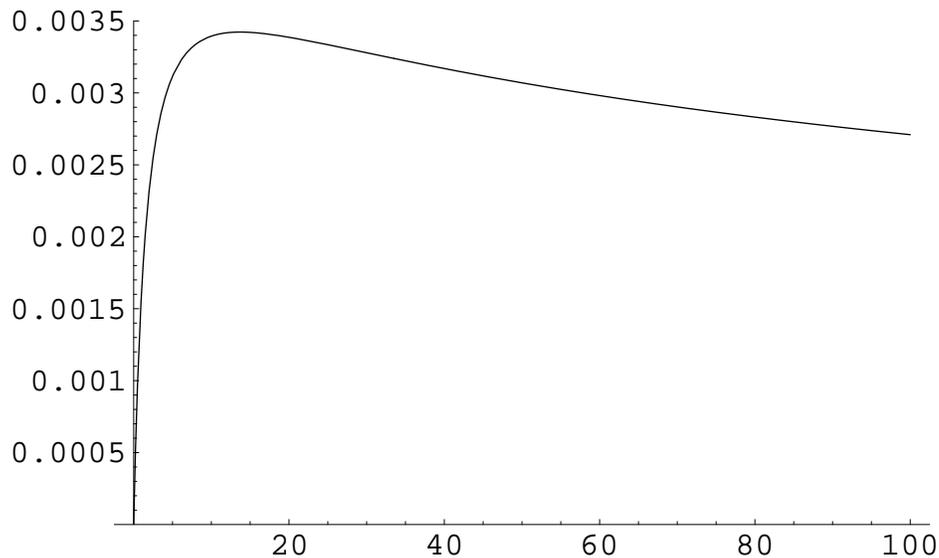}
\end{center}
\caption{The dependence of the expectation of the energy density on the
susceptibility.  The coefficient $\eta _\rho$ is plotted as a function of the
susceptibility $\chi$.  There is a single maximimum at $\chi \approx
13.65$.}
\end{figure}

The expectation of the energy density has the following limiting behaviours:\\
for $\chi \ll 1$ we have
\begin{equation}
\langle \hat{T}_{00} \rangle_{\rm ren} = \frac{\hbar c}{(2 \pi)^2 z^4}\left(
\frac{\chi}{10}-\frac{2 \chi^2}{35}
\right)+O(\chi^3)\, ,
\end{equation}
and for $\epsilon \gg 1$ we have
\begin{equation}
\langle \hat{T}_{00} \rangle_{\rm ren} = \frac{\hbar c}{(2 \pi)^2 z^4}
\frac{3}{8}\frac{\ln \epsilon}{\sqrt{\epsilon}}
+O(1/ \sqrt{\epsilon})\, .
\end{equation}

The graph of the coefficient $\eta _\rho$ is given in figure~2.  It has a
single maximum with value $\eta _\rho\approx .0034$ at $\chi\approx 13.65$. 
The fall--off is again very slow, with $\eta _\rho$ still at 50\%  its maximum
value for $\chi\approx 600$, and still at 10\%  its maximum for
$\chi\approx 60\,000$.

\section{Discussion}
We have found that the expectations of the renormalized
squares of the components of
the electric and magnetic fields at distance $z$ from a non--dispersive
dielectric all have the form $\eta\hbar c/z^4$, where the coefficents
$\eta$ depend on the susceptibility $\chi$.  These coefficients vary
linearly near $\chi =0$, and tend monotonically but very slowly to
asymptotic constant values as $\chi\uparrow\infty$.  For the electric field
components, one has $\eta >0$; while for the magnetic field one
has $\eta <0$.  For the energy density, the corresponding coefficient
$\eta _\rho$
rises linearly near $\chi =0$, attains a maximum at $\chi\approx 13.65$,
and then falls very slowly to zero as $\chi\uparrow\infty$.

Previous authors have considered the quantum optical effects of a
non--dispersive dielectric half--space on atomic transition rates 
(e.g. Khosravi \& Loudon 1991, 1992). 
Those results are in some sense complementary to these:  those test the
field operators over a small range of frequencies; whereas the present
ones depend on integrating over all frequencies.  The present results
depend crucially on the successful ultraviolet renormalization of the
theory.

\subsection{Casimir--Polder Forces}
An immediate consequence of these formulas is a prediction for the
Casimir--Polder force on a polarizable atom near a dielectric. 
If the polarizability is $\alpha (\omega )$, then the induced dipole is
$\int\alpha (\omega ){\hat E}(\omega ) \, \rmd\omega$, and the potential
energy is $-\left(\int\alpha (\omega ){\hat E}(\omega ) \,
\rmd\omega\right)\cdot {\hat E}$.  If in the regime in question we may
neglect the frequencey--dependence of the polarizability, then, in the
vacuum state of the electromagnetic field the potential energy is
the vacuum state of the electromagnetic field, this becomes
\begin{equation}
  -\alpha\langle{\hat E}^2\rangle _{\rm ren}=-\alpha\hbar c\eta ^E/z^4\,
,
\end{equation}
where $\eta ^E=\eta ^E_z+\eta ^E_{\rm T}$.  In principle, a parallel
treatment applies to derive a `magnetic Casimir--Polder' force
depending on magnetic polarizability.

The Casimir--Polder force on a polarizable atom near a conductor has
been measured in recent years (Sukenik {\it et al.} 1993),  although as
yet there has been no measurement near a dielectric.

\subsection{Relation to Consitutive Energy}
One of our main motivations for studying this model was to uncover its
limitations.  We noted early on that a real physical dielectric cannot be
well approximated as a uniform medium on arbitrarily small scales. 
This means that we cannot expect our model to accurately capture the
physics of the field modes whose wavelengths are less
than the atomic scale (certainly) or the skin depth (probably).  In
particular, our predictions must break down as one gets within a
distance of this order of the interface.

One vivid manifestation of this is the electromagnetic field energy of
the vacuum half--space, per unit surface area.  This surface energy
density is
\begin{equation}
 \int _0^\infty (\eta _\rho\hbar c/z^4)\, \rmd z\, ,
\end{equation}
which is {\em divergent} at the lower limit.  
(For an earlier investigation of a closely
related effect, see Bordag and Lindig 1996.)
It is not physically
plausible that this energy density diverges; rather, the divergence reflects
an improper model of physics near the interface.  A more correct version
would have the form
\begin{equation}
  \mbox{local contribution near interface} +\int _\delta ^\infty
(\eta _\rho \hbar c/z^4)\, \rmd z\, ,\label{hector}
\end{equation}
where $\delta$ is of the order of the skin depth.

The energies we are considering here represent electromagnetic
contributions to the constitutive energy of the medium.  (In a more
realistic treatment, it might not be meaningful to isolate one
class of electromagnetic contributions from others, however.)  It is
plausible that these depend very much on the chemical physics of the
material, and so the vagueness in the form (\ref{hector}) is apt.
The numerical value of
the contribution from the second term in (\ref{hector}) is quite modest
for everyday materials.  Taking the rather small value $\delta\approx
10\AA$ and $\eta _\rho =.003$, we find $\eta _\rho\hbar c/(3\delta
^3)\sim 3\cdot 10^{-4}\mbox{cal} /{\mbox{cm}}^2$.

\subsection{The Fields in the Dielectric}
In this paper, we have only treated the electromagnetic field in the
vacuum half--space outside the dielectric.  This case would seem to be
of more interest than the field within the dielectric.  Still, it is
natural to ask what would happen there.

In principle, techniques like ours should apply to compute the operators
${\hat D}=\epsilon{\hat E}$ and $\hat B$ within the dielectric.  The
integrals involved are more difficult than those on the vacuum side,
though.

Aside from technical difficulties in evaluating the integrals, there is
another point which must be considered in the dielectric region.  In
that region, the ultraviolet asymptotics of the two--point functions are
different than in Minkowski space.  (Because the two--point functions
are singular on characteristics, and there is a different speed of light
in the dielectric medium.)  This means that the renormalization cannot
be accomplished by subtracting the Minkowski--space vacuum quantities. 
One could presumably renormalize by subtracting the quantities
associated to a uniform dielectric.  If one does this, then the local
energy density differs from that of Minkowski space by an infinite
amount.

We believe the resolution to this point is the same as that discussed in
the previous subsection.  One cannot accept the present model as an
accurate picture of physics at all scales, and it is really only to be
considered as an effective field theory, valid for frequencies below
some cut--off.  The difference in energy densities should be finite,
with one contribution due to the effective field theory with a cut--off,
and another due to the details of the chemical physics of the medium.

\subsection{Sign of the Energy Density}
One of our motivations for this work was to get a better understanding
of the negative energy density occuring in the Casimir effect.  

The Casimir effect --- corresponding to two parallel plane conductors
--- is {\em formally} the limit as $\epsilon\uparrow\infty$ of two
parallel dielectric half--spaces.  However, this identification only
holds at the limit, and only in the sense that in this case the
reflections and refractions of the field modes at the dielectric
interfaces approach the perfect--conductor boundary conditions as
$\epsilon\uparrow \infty$.  A real conductor, with finite conductivity,
has a dielectric function which is significantly dispersive and
absorptive, and cannot be modeled by a constant large real positive
$\epsilon$.  Thus our present model cannot make any positive
quantitative predictions about
the Casimir effect for conductors.  

However, we shall show that at least
for dielectrics, the effects of a finite $\epsilon$ cannot be ignored,
and that in realistic situations it seems most likely that the total
expected energy density, including separation--independent contributions
from the half--spaces, is {\em positive}.
This suggests strongly that we must investigate the real physics of
conductors before we can conclude that the total expected energy density
between the plates is negative.

Consider two non--dispersive half--space 
dielectrics, of the same susceptibility,
parallel and separated by a distance $l$.  Then the total energy (per
unit cross--sectional area) will have the form
\begin{equation}
E_{\rm tot}=E_\infty +\eta _2\hbar c/l^3\, .
\end{equation}
The function $\eta _2(\chi )$ has been computed by Lifshitz (1956).  Here
$E_\infty$ is (twice) the energy of either dielectric in isolation.

For the energy density, let
$z$ with $-l/2 <z<l/2$ be a coordinate normal to the interface
planes.  Then near either interface one expects the energy density to be
dominated by the physics of that interface, and so to be $\sim\eta _\rho\hbar
c/(z\pm l/2)^4$, where $\eta _\rho(\chi )$ is the coefficient we computed
previously, equation (\ref{ichabod}).
We shall write these two contributions as $\rho _1(z)$
and $\rho _2(z)$.  The total energy density will be
\begin{equation}
\rho _{\rm tot} (z)=\rho _1 (z)+\rho _2(z) +\rho _{\rm Cas}(z,l)
\, ,
\end{equation}
where $\rho _{\rm Cas}(z,l)$ must on dimensional grounds have the form
$f(z/l)/l^4$, and be less singular at the interfaces than $\rho _1$,
$\rho _2$.  Indeed, the integral $\int _{-l/2}^{l/2} f(z/l) \rmd
z =\eta _2 l/(\hbar c)$ must be finite.  
Thus $f$ has at most mild singularities at the
interfaces.

The form of the function $f$ is at present unknown.  As a very rough
approximation, we shall assume it is constant in $z/l$, that is, the
`Casimir' contribution to the expected energy density is uniformly
distributed between the half--spaces.  We may ask if this value
dominates the contribution $\rho _1+\rho _2$ from the dielectrics,
that is, if $\rho _{\rm tot}$ is positive or negative. 
This question can be
answered by comparing our results with those of Lifshitz (1956),
who computed the force of attraction of the two dielectrics.  We find
numerically that the energy density $\rho_1 +\rho _2$ dominates the
average energy density unless $\chi\gtrsim 39\,000$.  In other words, if
the `Casimir' contribution to the energy density is distributed
uniformly, the total energy density would be everywhere positive between
the half--spaces unless $\chi$ could be made to exceed $\approx 39\,000$.

A real dielectric exhibits absorption and dispersion; we can
expect our model to be valid at distances of order $z$ if the
dielectric susceptibility is (nearly) a real positive constant for
several orders of magnitude of frequency bracketing $c/z$.  There
seems to be nothing in the Kramers--Kronig relations preventing this
from holding for the sorts of values of $\chi$ discussed above.
Still, the scales are extreme enough that one wonders whether such
susceptibility functions are more mathematical
curiosities than physical possibilities.  In other words, unless
remarkable materials exist, with $\chi (\omega )$ approximately a real
positive constant $\gtrsim 39\,000$ for several orders of magnitude of
$\omega$, it seems unlikely the total expected energy density anywhere
between the dielectric half--spaces will be negative.

We should like to emphasize that while the contributions from the
dielectrics probably make the energy density positive, their
separation--independence ensures that they do not alter the usual
predictions of the attractive force between the dielectrics.  Indeed, we
have used Lifshitz's results in our argument, and we accept his values
for the force.

\section{Conclusion}
Negative energy densities were first discovered in quantum field theory
with Casimir's prediction of an attractive force between two parallel
perfect plane conductors.  Since then, there has been considerable
speculation on what the physical consequences of these negative energy
densities might be.

The present model was introduced as a first step away from the
idealization of boundary conditions induced by a dielectric of infinite
susceptibility.  It has been chosen for its relative mathematical
simplicity, and it is unrealistic in that it neglects dispersion and
absorption.  Still, we find that the effects of finite--susceptibility
contributions go as $\epsilon ^{-1/2}\ln\epsilon$ and
can be very significant:  in the range that our model is
likely to be valid, it seems that these contributions can make the {\em
total} energy density between two dielectric half--spaces positive,
while preserving the attractive force found by Lifshitz.

One cannot, from our model, draw any definite conclusion about the
behavior of the energy density between two real conducting plates.
What sorts of separation--independent corrections there are, due to
finite conductivity, are at present unknown.  But it does seem clear
that we will only be justified in having confidence in theoretical
predictions of the energy density between two real conducting plates if
we take into account finite--conductivity effects.

\References
\item[] Barton G 1990 \PL {\bf B237} 559
\item[] Bordag M and Lindig J \JPA {\bf 29} 4481
\item[] Bordag M, Geyer B, Klimchitskaya G L and Mostepanenko V M 1998
\PR {\bf D58} 075003
\item[] Casimir H B G 1948 {\it Proc. K. Ned. Akad. Wet.} {\bf 51} 793
\item[] Helfer A D 1998 \CQG {\bf 15} 1169
\item[] Khosravi H and Loudon R 1991 \PRS {\bf A433} 337
\item[] \dash 1992 \PRS {\bf A436} 373
\item[] Lamoreaux S K 1997 \PRL {\bf 78} 5
\item[] Lifshitz E M 1956 {\it Sov. Phys. JETP} 
{\bf 2} {73 (Russian reference: 1955 {\bf
29} 94)}
\item[] L\"{u}tken C A and Ravndal F 1983 \PS {\bf 28} 209 
\item[] Matloob R, Loudon R, Barnett, S M and Jeffers, J 1995 \PR {\bf
A52} 4823
\item[] Milonni P W and Shih M 1992 {\it Contemporary Physics} {\bf 33}
313
\item[] Sparnaay M J 1957 {\it Nature} {\bf 180} 334
\item[] \dash 1958 {\it Physica} {\bf 24} 751
\item[] Sukenik C I, Boshier M G, Cho D, Sandoghdar V and Hinds E A 1993
\PRL {\bf 70} 560
\endrefs
\end{document}